# Revisiting the Plasmon Radiation Damping of Gold Nanorods

*Yanhe Yang[1+], Weihai Ni[1+], Hao Xie[1+], Jian You[1], and Weixiang Ye[1,2*]*

[1]School of Physical Science and Technology, Soochow University, Suzhou 215006, China
[2]Department of Applied Physics, School of Science, Hainan University, Haikou 570228, China

[+] contributed equally
[*]wxy@hainanu.edu.cn

Noble metal nanoparticles have been utilized for a vast amount of optical applications. For the applications that used metal nanoparticles as nanosensors and optical labeling, larger radiation damping is preferred (higher optical signal). To get a deeper knowledge about the radiation damping of noble metal nanoparticles, we used gold nanorods with different geometry factors (aspect ratios) as the model system to study. We investigated theoretically how the radiation damping of a nanorod depends on the material, and shape of the particle. Surprisingly, a simple analytical equation describes radiation damping very accurately and allow to disentangle the maximal radiation damping parameter for gold nanorod with resonance energy $E_{res}$ around 1.81 eV (685 nm). We found very good agreement with theoretical predictions and experimental data obtained by single-particle spectroscopy. Our results and approaches may pave the way for designing and optimizing gold nanostructure with higher optical signal and better sensing performance.

**TOC GRAPHICS:**

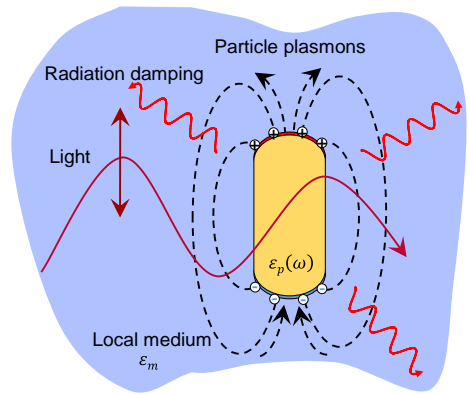

Noble metal nanoparticles have very large optical absorption and scattering cross-section due to resonantly driven electron oscillations upon interaction with light, which is also known as particle plasmons.[1] Particle plasmons enable the nanoparticles to collect light energy efficiently and stored it in the oscillation of the conduction electrons for tens of femtosecond. Then, the plasmon energy is redistributed via several channels: interband, intraband, radiation, and surface damping.[2-5] Among these, the radiation damping (radiative energy loss of particle plasmons) quantifies the efficiency of the transformation of the plasmon oscillation energy into the emitted photons. It is well known that radiation damping scales with the volume of the nanoparticles and plays a dominant part in the overall damping of particle plasmons for large metal nanoparticles.[6,7] For the applications that used metal nanoparticles as nanosensors and optical labeling, larger radiation damping is preferred (higher optical signal). [8-12]

Historically, radiation damping is related to the volume of the nanoparticles, $V$, and the radiation damping parameter, $\kappa_{rad}$, according to $\Gamma_{rad} = 2\hbar\kappa_{rad}V$. [6] However, we noticed that such description may have deviations for particles with different geometry factors. For example, for gold nanorods with different aspect ratios. [6,7,13] We have therefore revisited this problem and derived an analytical equation that accurately describes the radiation damping for gold nanorods. The derivation is based on the fundamental light scattering equation from Clausius-Mossotti which uses the complex susceptibilities of the particle material and geometry factors as input.[14] Applying the theory to gold nanorods with different aspect ratios, we show theoretically and with experiments how the historical radiation damping parameter depends on the aspect ratio. Our results indicate that there is an optimal radiation damping parameter for gold nanorod with a certain aspect ratio. Such gold nanorod has plasmon resonance which corresponds to the minimal imaginary part of the dielectric function of gold. Intuitively, the minimal imaginary part of the dielectric function for a material corresponds to the minimal energy loss. Our results and approaches may pave the way for designing and optimizing gold nanostructure with higher optical signal and better sensing performance.

In this letter, we have chosen gold nanorods with different geometry factors (aspect ratios) as the model system to study. There are two reasons for this: recipes are available for making gold nanorods with good control of sizes and aspect ratios, and they have dipole plasmon resonances in the visible light region of the spectrum which can be easily studied.[15] As mentioned above, one of the plasmon energy decay channels is radiative decay. The gold nanorod is able to transform the collective electron oscillation energy into the optical far-field under light excitation (cf. FIG.1.a). In order to determine the radiation damping from the homogeneous plasmon line width (the overall plasmon damping) of the gold nanorods with high statistics, we used the spectral imaging single-particle darkfield spectroscopy to collect the scattering spectrum of the particles. As shown in FIG.1.b, we used an upright microscope equipped with a darkfield condenser and a liquid crystal tunable filter as our spectral imaging setup.[11] The scattered light of the gold nanorods under different wavelengths of excitation are collected by a CMOS camera. Such configuration allows us to record all the particles within the field of view (FOV) simultaneously. From the recorded scattering intensities at different wavelengths, we can obtain the plasmon resonance energy ($E_{res}$) and linewidth ($\Gamma$) with a Lorentzian function (cf. FIG.1.c).

The particles investigated here were produced chemically. The gold nanorods have diameters from 29 to 92 nm and aspect ratios from 1.6 to 2.8 (cf. FIG.S1 and Table.S1). Please note that the surface damping gives negligible contributions to the total plasmon damping for the particles used in this work.[16] We have used the scanning electron microscope (SEM) and single-particle spectroscopy to correlate the mean volume and mean radiation damping of the same batch of gold nanorods. To achieve this, we first determined the diameter ($D$) and length ($L$) of thousands of nanorods from the SEM images (cf. FIG.2.a and FIG.2.b). Then, we calculated the particle volume ($V$) with these values ($D$, $L$) by modeling the gold nanorods as a spherically capped cylinder shape for every particle. The mean volume ($<V>$) of one batch of the gold nanorods can be obtained by fitting the distribution of the nanoparticles' volume with a gaussian function (cf. FIG.2.c). Following, we measured the scattering spectrum of the same batch of gold nanorods with our spectral imaging setup and determined the total plasmon damping ($\Gamma$) for every particle (cf. FIG.2.d). It should be noted that the $\Gamma$ depends on the plasmon resonance energy ($E_{res}$) due to the contributions of the bulk damping (interband and intraband damping) as nonradiative damping $\Gamma_{non-rad}$. The bulk damping of the dipole plasma resonance can be estimated by using the real part $\varepsilon_1$ and imaginary part $\varepsilon_2$ of the metal-dielectric function as:[17]

$$\Gamma_{non-rad} = 2\varepsilon_2/|\varepsilon_1'| \qquad (1)$$

The $\Gamma_{non-rad}$ also depends on the $E_{res}$ and shows an increasing trend for higher energy due to the contribution of interband damping (the dash line in FIG.2.d). We can extract the radiation damping of each gold nanorod by subtracting the contribution of the bulk damping from total plasmon damping the as $\Gamma_{rad} = \Gamma - \Gamma_{non-rad}$. It is clear that the radiation damping does not have a strong dependency on the $E_{res}$ in a narrow energy range (cf. FIG.2.e). The mean radiation damping <$\Gamma_{rad}$> of the same batch of the gold nanorods can be obtained by fitting the distribution of the $\Gamma_{rad}$ for thousand nanoparticles with a gaussian function (cf. FIG.2.f). Finally, we are able to correlate the <$\Gamma_{rad}$> to a certain <$V$> of the same batch of gold nanorods. We have repeated this procedure for all the batches of gold nanorods used in this work.

Before introducing our analytical equation for radiation damping, we have determined the mean radiation damping parameter <$\kappa_{rad}$> from <$\kappa_{rad}$>= <$\Gamma_{rad}$>/ $2\hbar$ <$V$> for all batches of gold nanorods. Our experimental data follows very well to this equation (cf. FIG.3.a) and the value <$\kappa_{rad}$>= $6.55 \pm 0.02 \cdot 10^{-7}$ fs$^{-1}$ nm$^{-3}$ is in the same order as the previously reported values.[6,7] However, this equation overlooks the contribution of the geometry factors of the particles to the radiation damping. In our case, for gold nanorods, is the contribution of aspect ratios to the radiation damping. Therefore, we have revisited this problem and derived an analytical equation based on the fundamental light scattering equation from Clausius-Mossotti to describes the radiation damping for gold nanorods. The plasmon resonance of a nanoparticle is the frequency or wavelength where the polarizability $\alpha$ reaches a maximum. For particles with dimensions smaller than the wavelength of light, this polarizability is given by the Clausius-Mossotti equation (in the form of Richard Gans):[18]

$$\alpha = V \frac{\varepsilon_p - \varepsilon_m}{\varepsilon_m + L(\varepsilon_p - \varepsilon_m)} \quad (2)$$

Here, $V$ is the particle volume, $L$ is a geometry factor that is approximately $\frac{1}{L} \approx (1 + AR)^{1.6}$ for nanorods [12], $\varepsilon_m$ is the dielectric function of the surrounding medium (in our case, is water) and $\varepsilon_p = \varepsilon_1 + i\varepsilon_2$ is the particles' complex susceptibility (dielectric function) of gold. In metals, the real part of the particles' susceptibility $\varepsilon_1$ can be negative and, especially for coin and alkali metals, the imaginary part $\varepsilon_2$ comparatively small. As the dielectric function of the surrounding medium usually is positive, this leads to a strong maximum of polarizability at specific resonance energy ($E_{res}$) or resonance wavelength ($\lambda_{res}$). If $\varepsilon_2$ depends only weakly on wavelength, the polarizability has a maximum near the place where the real part of the denominator in equation (2) is zero:

$$\Re(|\varepsilon_m + L(\varepsilon_p - \varepsilon_m)|) = 0 \quad (3)$$

Equation (3) determines the resonance condition for the nanoparticles in a certain surrounding medium:

$$\varepsilon_1 = \varepsilon_m \left(1 - \frac{1}{L}\right) \tag{4}$$

Combining equation (2) and equation (4), we can express the polarizability at resonance condition:

$$\alpha_{res} = \frac{V}{L}\left(1 + \frac{i\varepsilon_m}{L\varepsilon_2}\right) \tag{5}$$

The corresponding maximum cross sections for scattering and absorption $C_{sca}$ and $C_{abs}$ are:

$$C_{sca} = \frac{k^4}{6\pi}|\alpha|^2\bigg|_{at\ res} = \frac{k^4}{6\pi} \cdot \frac{V^2}{L^2}\left(1 + \frac{\varepsilon_m^2}{L^2\varepsilon_2^2}\right) \tag{6}$$

and

$$C_{abs} = k\Im(\alpha)|_{at\ res} = k\frac{V}{L^2} \cdot \frac{\varepsilon_m}{\varepsilon_2} \tag{7}$$

Here k is the wavevector $k = \frac{2\pi\sqrt{\varepsilon_m}}{\lambda_{res}}$ at resonance wavelength.

Combining equation (2), equation (6), and equation (7), we obtained the analytical equation for radiation damping with the knowledge of the value of nonradiative damping, $\Gamma_{non-rad}$, scattering cross-section, $C_{sca}$, and absorption cross-section, $C_{abs}$, at resonance condition:[19]

$$\Gamma_{rad} = \frac{C_{sca}}{C_{abs}}\bigg|_{at\ res} \cdot \Gamma_{non-rad} = \frac{4\pi^2}{3} \cdot \left(\frac{\sqrt{\varepsilon_m}}{\lambda_{res}}\right)^3 \cdot \frac{V}{L} \cdot \frac{L^2\varepsilon_2^2 + \varepsilon_m^2}{L\varepsilon_2\varepsilon_m} \cdot \frac{2\varepsilon_2}{|\varepsilon_1'|} \tag{8}$$

This equation can be further simplified by expressing the $\lambda_{res}$ in an analytical form with the help of the Drude model. In this case, the dielectric function of the particle can be written as:[1]

$$\varepsilon(\omega) \approx \varepsilon_\infty - \frac{\omega_p^2}{\omega^2} + i\frac{\omega_p^2\gamma}{\omega^3} \tag{9}$$

Here, $\varepsilon_\infty = 9.84\ eV$ is the background permittivity which used to correct the contribution of d-band electrons that are close to the Fermi surface, $\omega_p = \sqrt{Ne^2/\varepsilon_0 m^*} = 9\ eV$ is the bulk plasmon frequency with effective electron mass $m^*$ due to the band structure corrections, and $\gamma = 67$ meV is the damping of the free electrons.

To show how nicely the Drude model predicts the dielectric function of gold, we have plotted the real and imaginary parts of the dielectric function of bulk gold calculated using the above Drude parameters and experimentally data measured by Johnson and Christy (JC) in FIG.S2. With the real part of the Drude dielectric function $\varepsilon_1 = \varepsilon_\infty - \omega_p^2/\omega^2$ from equation (9) inserted into the resonance condition, equation (4), we can express the resonance wavelength as:

$$\lambda_{\text{res}} = 2\pi c_0/\omega_{res} = \lambda_p \sqrt{\varepsilon_\infty + \varepsilon_m(\frac{1}{L} - 1)} \quad (10)$$

Where $\lambda_p = 2\pi c_0/\omega_p$ is the bulk plasmon wavelength and $c_0$ is the speed of light in vacuum. Please noted that this is the eigen resonance wavelength of particle plasmons. We have plotted resonance wavelengths of gold ellipsoidal nanoparticles predicted by equation (10) and calculated by *quasi-static approximation* (QSA) simulation in FIG.S3. The results from the equation are matched surprisingly well with the QSA simulation.

Combining equation (10) and equation (8), we reached the final analytical equation of radiation damping:

$$\Gamma_{rad} = \frac{8\pi^2}{3} \cdot \frac{1}{\lambda_p^3(\varepsilon_\infty - \varepsilon_1)^{3/2}} \cdot \left(\varepsilon_2^2 + \frac{\varepsilon_1^2}{(L-1)^2}\right) \sqrt{\frac{L\varepsilon_1}{(L-1)(|\varepsilon_1'|)^2}} \cdot V \quad (11)$$

It is clear that the radiation damping of a plasmonic nanoparticle depends on the geometry factor of the particle $L$, the bulk plasmon wavelength $\lambda_p$, the dielectric function and the material background permittivity $\varepsilon_\infty$ of the material. Comparing our analytical equation with the historical expression $\Gamma_{rad} = 2\hbar\kappa_{rad}V$, we can obtain the analytical expression for the radiation damping parameter:

$$\kappa_{rad} = \frac{4\pi^2}{3\hbar} \cdot \frac{1}{\lambda_p^3(\varepsilon_\infty - \varepsilon_1)^{3/2}} \cdot \left(\varepsilon_2^2 + \frac{\varepsilon_1^2}{(L-1)^2}\right) \sqrt{\frac{L\varepsilon_1}{(L-1)(|\varepsilon_1'|)^2}} \quad (12)$$

To show how precisely the analytical equation of radiation damping for gold nanorods, we have plotted our mathematical predictions of the radiation damping of gold nanorods (green cross) by using the mean volume and mean aspect ratio from SEM images as input values, and the experimental data obtained from single-particle spectroscopy in FIG.3.b. The error bars for the radiation damping and volume of the particles have been omitted to avoid cluttering of the figure. The deviations of our mathematical predictions to the experimental data are below 20% for the different sizes of gold nanorods by using the experimental data as reference (cf. FIG.3.c). We have noticed that the radiation damping parameter is a function of geometry factor from the equation (12), which means the particles with different geometry factors (In our case, is aspect ratio) have different efficiency to transform the collective electron oscillation energy into the optical far-field. Therefore, we have plotted the radiation damping parameter as a function of nanorods' aspect ratio (cf. FIG.3.d). The blue dots and green crosses are experimental data and our mathematical predictions, respectively. The black line is the extrapolation by using gold nanorods with a constant diameter (20 nm) but a different aspect ratio (1.5 to 10). The experimental data and our mathematical predictions follow the same dependency on aspect ratio, which is first increased and then decreased. Interestingly, our mathematical prediction indicates that there is a maximal radiation damping parameter for gold nanorod with an aspect ratio of around 3.15. Such gold nanorod has $E_{res}$ around 1.81 eV (685 nm) which corresponds to the minimal point of the imaginary part of the dielectric function of bulk gold (cf. FIG.S.2). In the picture of particle plasmon, this maximal point for radiation damping parameter may correspond to the minimal point of nonradiative damping, $\Gamma_{non-rad}$, due to the suppression of interband damping. This maximal radiation damping parameter should make gold nanorods with $E_{res}$ around 1.81 eV becomes one of the best candidates in optical applications where large optical far-field signal is required, such as single-particle biosensing.

In conclusion, we have investigated the radiation damping of particle plasmons in single gold nanorods and derived an analytical formula describing the radiation damping. We have shown that experimental data of radiation damping follows quite well the trend predicted by our theory. We have found a maximal radiation damping parameter for gold nanorod with resonance energy $E_{res}$ around 1.81 eV (685 nm), which corresponds to the minimal point of the imaginary part of the dielectric function of gold. This finding indicates that gold nanorods with such resonance energy have relatively high light-scattering efficiencies and large optical far-field signals. A major advantage of the analytical equation of radiation damping is that it can be used to compare different nanoparticles: plasmonic particles of different materials and shapes, such as silver nanorods and silver spheres. We believe that our results and approaches could pave the way for designing and optimizing gold nanostructure with higher optical signal and better sensing performance.

ASSOCIATED CONTENT

**Supporting Information.** Supplementary text describing the synthesis method, the morphology and optical characterization of gold nanorods, and additional figures showing the

Drude model for the real and imaginary parts of the dielectric function of gold, the resonance wavelengths ($\lambda_{res}$) of gold nanorods predicted from equation (10) with Drude parameters.

AUTHOR INFORMATION

The authors declare no competing financial interest.

**Corresponding Author**

Weixiang Ye, wxy@hainanu.edu.cn

**Author Contributions.** W.Y. initiated and designed the research. W.Y. derived the mathematical models, constructed the microscope setup, gave guidance on the optical measurement, interpreted the data, performed data analysis, and written the manuscript text. Y.Y. and H.X performed the single-particle characterization and SEM measurements. W. N. provided the chemical resources and instruments for gold nanorods synthesis. Y.Y., H.X, and J.Y. performed the gold nanorods synthesis and size characterization.

ACKNOWLEDGMENT

W.Y. acknowledges the financial support by The Natural Science Foundation of Jiangsu Province (BK20200875) and The Natural Science Foundation of The Jiangsu Higher Education Institutions of China (20KJD150003). W.Y. thank Prof. Dr. Carsten Sönnichsen from the University of Mainz for constructive discussions about particle plasmons during the last six years.

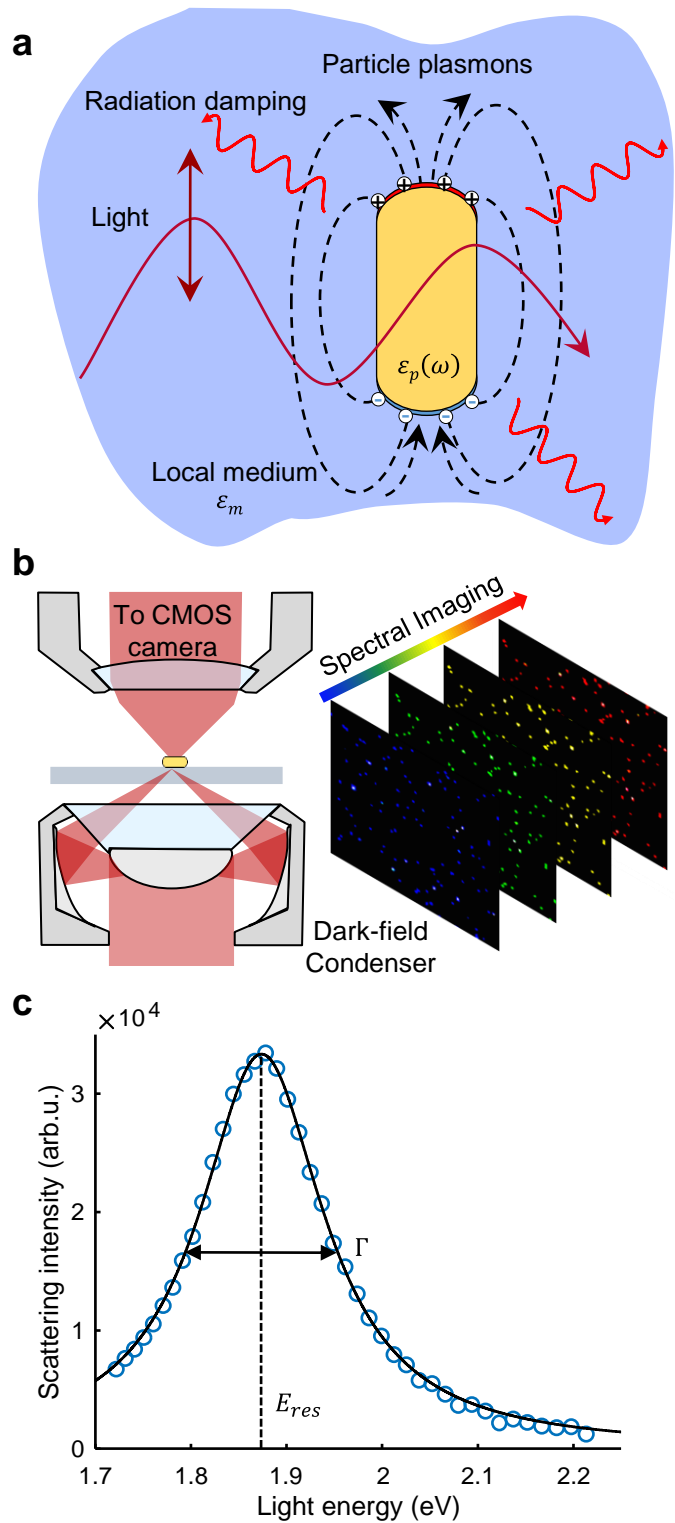

**FIG.1. Radiative decay of particle plasmons and single-particle darkfield scattering spectroscopy.** (a) Schematic representation of the radiative decay of particle plasmons in gold nanorod. The metal nanoparticle behaves as an optical dipole antenna under light excitation, which is able to transform the collective electron oscillation energy into the optical far-field. (b) The scattering spectrum of the gold nanorods can be obtained by using single-particle darkfield scattering spectroscopy. Here, we used the spectral imaging method to record the scattering spectrum of thousands of particles in the field of view simultaneously. (c) We can quantify the plasmon resonance energy ($E_{res}$) and linewidth ($\Gamma$) by fitting the recorded spectrum with a Lorentzian function.

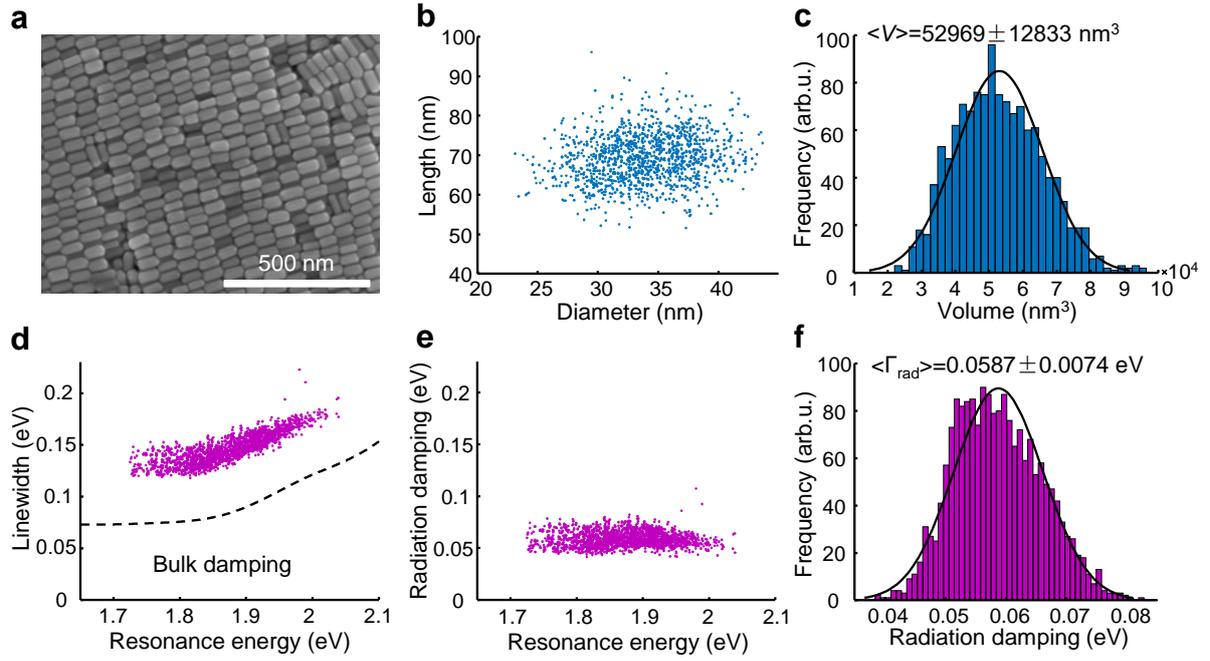

**FIG.2. We used the scanning electron microscope (SEM) and single-particle spectroscopy to correlate the mean volume and mean radiation damping of the same batch of gold nanorods**. (a) Representative scanning electron microscopy (SEM) image of the gold nanorods used in this work. (b) We obtained the diameter ($D$) and length ($L$) of thousands of nanorods from the SEM images. (c) We calculated the particle volume ($V$) with these values($D$, $L$) by modeling the gold nanorods as a spherically capped cylinder shape. The histogram shows the distribution of the nanoparticles' volume for around one thousand nanoparticles. (d) We measured the plasmon linewidth ($\Gamma$) of the same batch of gold nanorods with the single-particle darkfield scattering spectroscopy operating in the spectral imaging mode. Each pink dot represents the data obtained from one single nanoparticle. It is clear that $\Gamma$ depends on the plasmon resonance energy ($E_{res}$), which follows a similar trend of bulk damping. (e) We extracted the radiation damping ($\Gamma_{rad}$) of each gold nanorod by subtracting the contribution of the bulk damping. The radiation damping does not have a strong dependency on the $E_{res}$. (f) The histogram showing the distribution of the $\Gamma_{rad}$ for around two thousand nanoparticles.

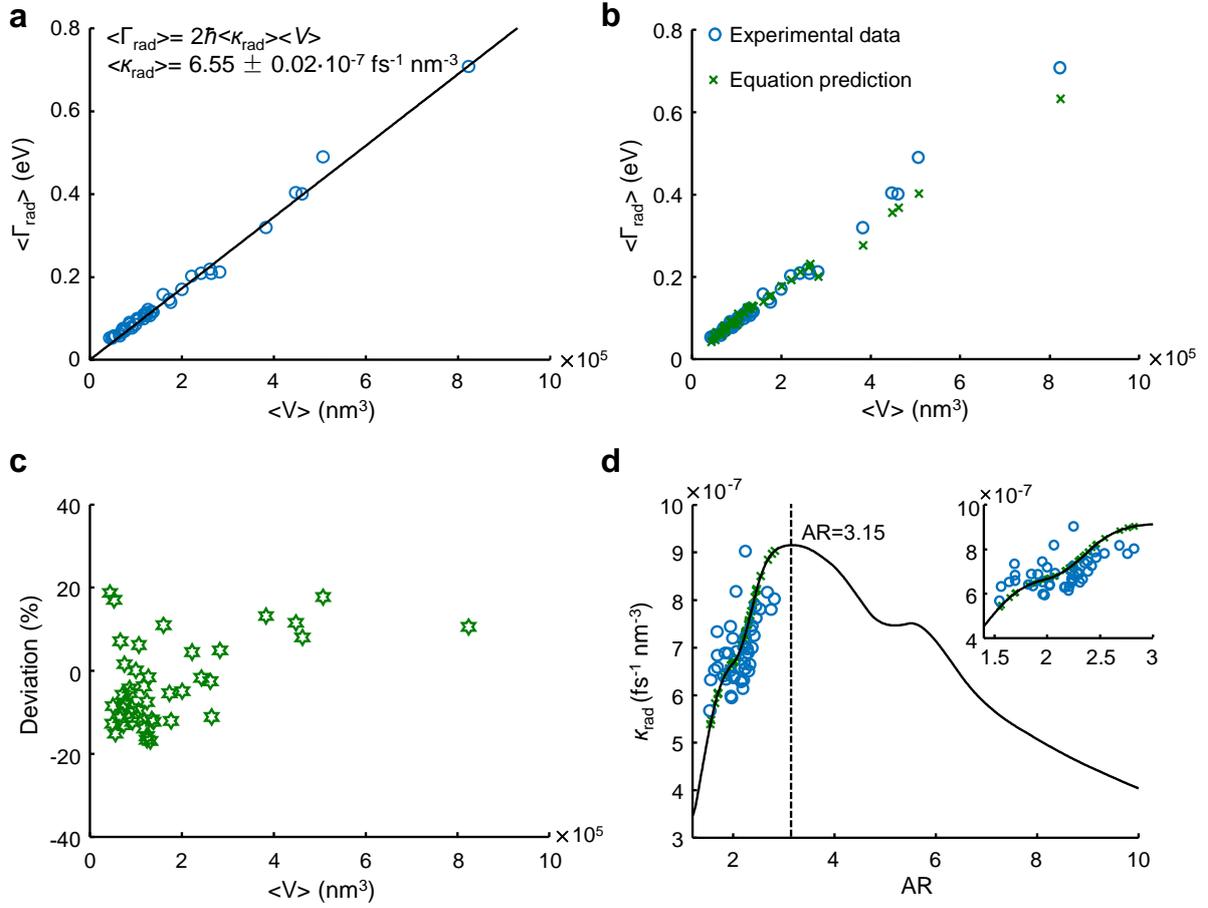

**FIG.3. Comparison of experimental data to our mathematical description for radiation damping of gold nanorods**. (a) The radiation damping ($\Gamma_{rad}$) of gold nanorods follows a quasi-linear trend with the particle volume (blue dot). We determined the mean radiation damping parameter from the best linear fit. (b) Our mathematical predictions of the radiation damping of gold nanorods (green cross) follow the same trend as the experimental data by using the mean volume and mean aspect ratio from SEM images as input values. (c) The deviations of our mathematical predictions to the experimental data are below 20% for the different sizes of gold nanorods. (d) The radiation damping parameter is a function of particles aspect ratio. The blue dots and green crosses are experimental data and our mathematical predictions. The black line is the extrapolation by using gold nanorods with a constant diameter (20 nm) but a different aspect ratio (1.5 to 10). The experimental data and our mathematical predictions show the same dependency on aspect ratio. Our mathematical predictions indicate that there is an optimal radiation damping parameter for gold nanorod with an aspect ratio of around 3.15. Such gold nanorod has $E_{res}$ around 1.81 eV (685 nm) which corresponds to the minimal point of the imaginary part of the dielectric function of bulk gold.

# Supporting Information for
## Revisiting the Plasmon Radiation Damping of Gold Nanorods

Yanhe Yang[1+], Weihai Ni[1+], Hao Xie[1+], Jian You[1], and Weixiang Ye[1,2*]

[1]School of Physical Science and Technology, Soochow University, Suzhou 215006, China
[2]Department of Applied Physics, School of Science, Hainan University, Haikou 570228, China

[+] contributed equally
[*]wxy@hainanu.edu.cn

**This PDF file includes:**
        Materials and Methods
        Figures S1 to S3 and Table S1.

## MATERIALS AND METHODS

**Materials**. All chemicals were obtained from commercial suppliers and used without further purification. Cetyltrimethylammonium bromide (CTAB, ≥98%) was purchased from Kermel. sodium oleate (NaOL, >97%) was purchased from TCI. Gold chloride trihydrate (HAuCl$_4$ · 3H$_2$O) was purchased from Aladdin. L-ascorbic acid (AA), Sodium borohydride (H$_4$BNa, ≥98%), silver nitrate (AgNO$_3$, ≥99%), Hydrochloric acid (HCl, 37 wt. % in water) were purchased from Sigma Aldrich. Ultrapure water obtained from a Direct-Q Water Purification System was used in all experiments.

**Synthesis and characterization of gold nanorods**. We have synthesized different gold nanorod batches to have a variety in aspect ratio and volume $V$ (cf. FIG.S1). The gold nanorods were prepared by a two-step seeded-growth process according to reference 1 with down-scale modifications. We have listed the details about the chemical amounts for each batch in Table S1. The dimensions of the synthesized gold nanorods were determined by scanning electron microscopy (SEM) using a Regulus 8100 microscope with a Cary 60 UV-Vis spectrophotometer. FIG.S1 shows the representative SEM images of the gold nanorods. From those images, we obtained the diameter ($D$) and length ($L$) of thousand of nanorods and calculated with these values the particle volume by $V = \frac{\pi}{6} D^3 \left(\frac{3L}{2D} - \frac{1}{2}\right)$, which is also listed in Table.S1.

Specifically, in the synthesis of gold nanorods, 0.28 g (0.037M in the final growth solution) or 0.36 g (0.047 M in the final growth solution) of CTAB and a certain quantity of NaOL were first dissolved in 10 ml of warm water (~50 °C) in a 50 ml centrifuge tube. The solution was allowed to cool down to 30 °C and 4mM AgNO3 solution was added. Following, 10 mL of 1 mM HAuCl4 solution was added and mixed thoroughly, then the mixture was kept distilled at 30 °C for 15 min. The solution became colorless after 90 min of stirring (SPH-103D, 300 rpm) and a certain volume of HCl (37 wt. % in water, 12.1 M) was added (Table.S1). After another 15 min of slow stirring (SPH-103D, 160 rpm), 0.05 mL of 0.064 M ascorbic acid (AA) was added and the solution was quickly stirred (SPH-103D, 200 rpm) for 30 s. Finally, a small volume of seed solution was injected into the growth solution (Table.S1). The resultant mixture was stirred (SPH-103D, 200 rpm) for 30 s and then left undisturbed was put into a water bath at 28 °C for 12 h for gold nanorods growth. We have removed the growth reagents from the gold nanorods' solution by centrifuging the gold nanorods' solution for 30 min with 3214 g and re-dispersing the sedimentation into 5 mL ultrapure water and 5 mL of 0.1 M CTAB solution.

**Single-particle spectroscopy.** We used a home-built spectral imaging single-particle darkfield microscope to determine the radiation damping from the homogeneous plasmon line width of the gold nanorods with high statistics. The microscope setup consists of an upright microscope Zeiss Axioscope 5 with an EC-Epiplan 50x/0.75 Zeiss objective, a Plan-Apochromat 63x/1.4 oil lris Zeiss objective, oil-immersion darkfield condenser (NA 1.2-1.4), a Halogen Lamp (OSL2, Thorlabs), and a liquid crystal tunable bandpass filter (KURIOS-VBI, Thorlabs). The scattered light of the gold nanorods under different wavelengths of excitation are collected by a COMS camera (Hamamastu orca flash V4.0). we used a self-written MATLAB software program to determine the plasmon resonance energy ($E_{res}$) and linewidth ($\Gamma$) from all the particles.

**Sample preparation for single-particle spectroscopy.** To get sparsely distributed gold nanorods for single-particle measurements, we centrifuged 1 mL of gold nanorods solution (in 0.05 M CTAB) twice and redispersed the sedimentation in 1.2 mL ultrapure water. Following, we immobilized the particle on a piece of clean glass coverslip (Deckgläser, 24 x 60 mm) by drop-casting a centrifuged solution of gold nanorods (~50 μL), and then we rinsed the cover glass with ultrapure and ethanol separately. After dring the cover glass, we placed a piece of clean glass coverslip (Deckgläser, 24 x 24 mm) on the center of the glass coverslip (Deckgläser, 24 x 60 mm) with gold nanorods. We used four the small glasses (Deckgläser, 2 x 2 mm) as gap holder, which were placed at the four corners of the cover glass to (Deckgläser, 24 x 24 mm) so that a gap of about 0.2 mm thick was formed between the two pieces of glass. Finally, The glass coverslip (Deckgläser, 24 x 24 mm) was fixed with tape (0.5 cm wide and 3.5 cm long) along the long side of the glass coverslip (Deckgläser, 24 x 60 mm), and then ultrapure water was injected to fill the interlayer, which allows us to measure the particle in water-immersion condition.

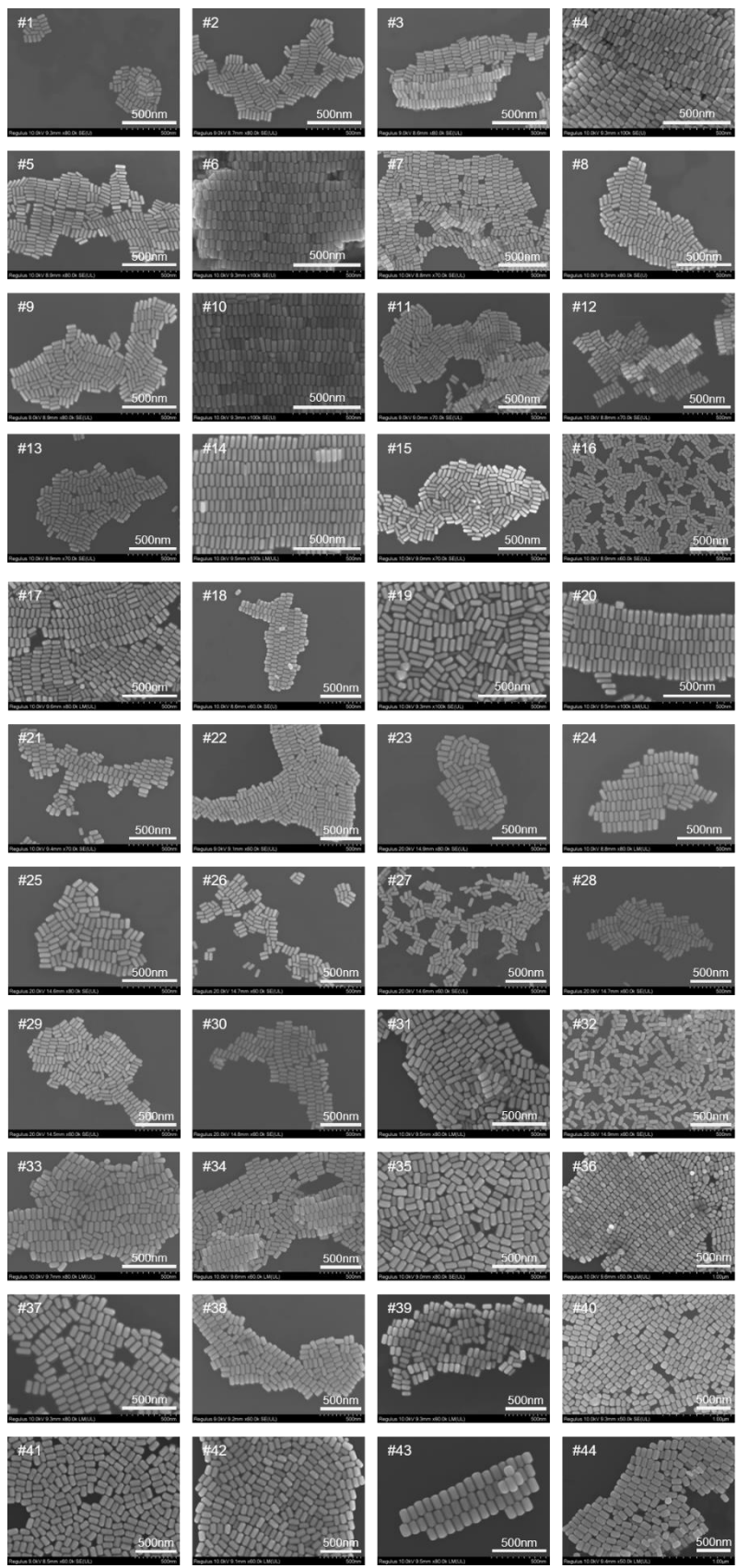

**FIG.S1. Representative SEM images of the 44 batches of gold nanorods used in this work.** The scale bar is 500 nm and the sample number is indicated by the white number. The particles are ordered by their volume.

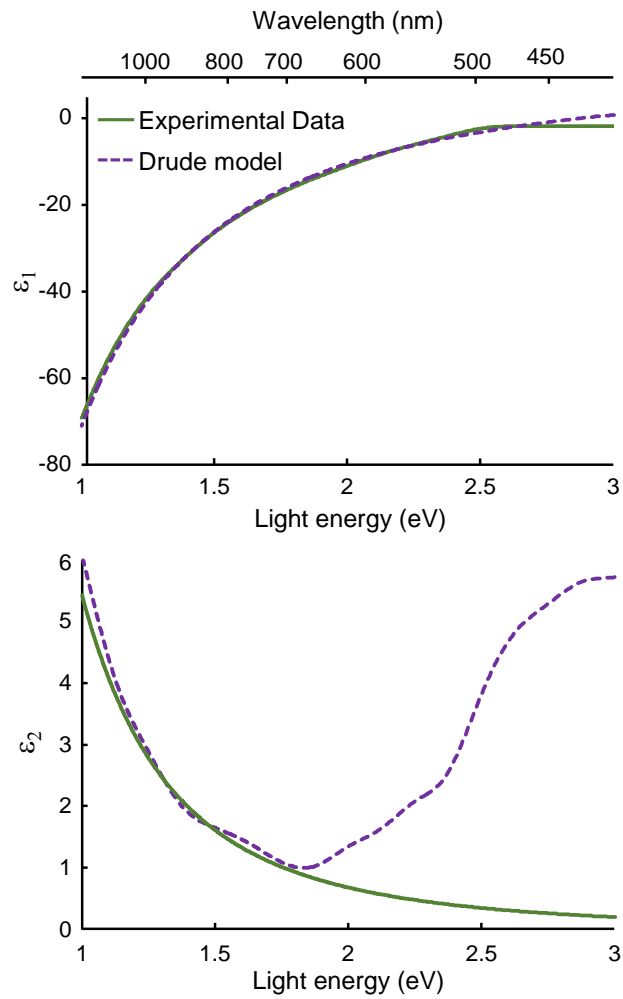

**FIG.S2. Experimental data and the Drude model for the real and imaginary parts of the dielectric function of bulk gold**. Experimental data (purple dot line) were measured by Johnson and Christy, the Drude model (green line) was calculated from the given parameters.

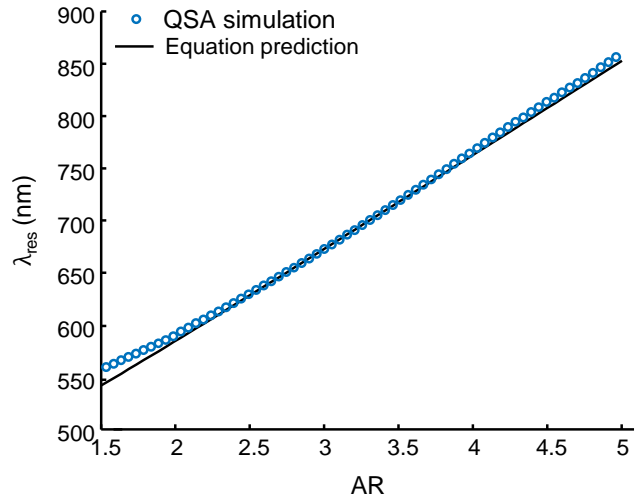

**FIG.S3. The resonance wavelengths ($\lambda_{res}$) of gold nanorods are predicted from equation 10 (line) with Drude parameters**. The QSA simulation (dots) is used for Comparison. In the case of QSA simulation, the data were calculated with gold ellipsoidal nanoparticles for 20 nm diameter and different aspect ratios from 1.5 to 5 in 0.05 steps.

| | CTAB (g) | AgNO$_3$ /4mM (ml) | Seed (ml) | HCl (ml) | NaOL (g) | Mean Length (nm) | Mean Diameter (nm) | Mean Volume (nm$^3$) | Aspect ratio |
|---|---|---|---|---|---|---|---|---|---|
| #1 | 0.36 | 0.4 | 0.08 | 0.06 | 0.06172 | 68.55±6.04 | 30.83±3.48 | 43928.70±10175.87 | 2.25±0.32 |
| #2 | 0.36 | 0.72 | 0.064 | 0.06 | 0.06172 | 81.85±8.25 | 29.37±3.17 | 49107.33±10854.51 | 2.82±0.45 |
| #3 | 0.36 | 0.72 | 0.08 | 0.06 | 0.06172 | 79.96±7.37 | 30.14±3.11 | 50113.35±10159.51 | 2.69±0.40 |
| #4 | 0.36 | 0.32 | 0.08 | 0.06 | 0.06172 | 69.28±6.10 | 33.91±3.83 | 52969.09±12833.35 | 2.07±0.28 |
| #5 | 0.36 | 0.72 | 0.08 | 0.06 | 0.06172 | 84.18±7.85 | 30.87±3.33 | 55652.37±11897.27 | 2.76±0.42 |
| #6 | 0.36 | 0.48 | 0.08 | 0.06 | 0.06172 | 78.61±7.15 | 35.41±3.31 | 66146.37±13081.82 | 2.24±0.30 |
| #7 | 0.36 | 0.36 | 0.08 | 0.06 | 0.06172 | 73.07±6.81 | 37.04±4.71 | 66372.81±17662.59 | 2.00±0.29 |
| #8 | 0.36 | 0.56 | 0.08 | 0.06 | 0.06172 | 83.89±7.67 | 35.49±3.20 | 71609.03±13794.35 | 2.38±0.32 |
| #9 | 0.36 | 0.72 | 0.048 | 0.06 | 0.06172 | 87.61±8.38 | 34.79±3.29 | 72567.79±14177.47 | 2.54±0.36 |
| #10 | 0.36 | 0.64 | 0.08 | 0.06 | 0.06172 | 84.62±6.97 | 35.75±3.34 | 73380.55±14189.88 | 2.39±0.30 |
| #11 | 0.36 | 0.72 | 0.032 | 0.06 | 0.06172 | 87.13±7.77 | 35.59±2.79 | 75041.66±12212.95 | 2.47±0.32 |
| #12 | 0.36 | 0.44 | 0.08 | 0.06 | 0.06172 | 78.45±6.88 | 38.13±3.67 | 75465.06±15194.10 | 2.08±0.28 |
| #13 | 0.36 | 0.52 | 0.08 | 0.06 | 0.06172 | 82.82±7.69 | 37.10±3.62 | 76449.47±14872.13 | 2.26±0.33 |
| #14 | 0.36 | 0.72 | 0.016 | 0.06 | 0.06172 | 91.16±7.04 | 37.38±3.07 | 86707.67±14581.14 | 2.46±0.29 |
| #15 | 0.36 | 0.68 | 0.08 | 0.06 | 0.06172 | 88.90±8.47 | 38.18±4.13 | 87812.06±19229.61 | 2.36±0.35 |
| #16 | 0.36 | 0.6 | 0.08 | 0.06 | 0.06172 | 87.46±7.81 | 39.59±3.83 | 91817.63±17906.88 | 2.23±0.31 |
| #17 | 0.36 | 0.48 | 0.096 | 0.04 | 0.06172 | 85.83±6.96 | 40.06±4.44 | 92350.45±21515.71 | 2.17±0.28 |
| #18 | 0.36 | 0.24 | 0.08 | 0.06 | 0.06172 | 79.58±7.61 | 44.05±4.69 | 100096.39±23785.88 | 1.82±0.24 |
| #19 | 0.28 | 0.72 | 0.096 | 0.06 | 0.04936 | 94.83±10.67 | 39.83±5.67 | 103953.99±30581.56 | 2.42±0.41 |
| #20 | 0.28 | 0.72 | 0.048 | 0.06 | 0.04936 | 93.35±8.27 | 40.89±4.17 | 105769.08±23283.46 | 2.30±0.29 |
| #21 | 0.36 | 0.28 | 0.08 | 0.06 | 0.06172 | 81.65±8.68 | 44.56±5.76 | 106577.12±31714.22 | 1.85±0.25 |
| #22 | 0.36 | 0.72 | 0.016 | 0.06 | 0.06172 | 95.17±7.85 | 43.04±3.84 | 117984.80±21251.92 | 2.23±0.29 |
| #23 | 0.36 | 0.64 | 0.08 | 0.06 | 0.06172 | 94.73±8.19 | 43.19±3.94 | 118406.99±22680.19 | 2.21±0.28 |
| #24 | 0.28 | 0.72 | 0.032 | 0.06 | 0.04936 | 99.34±8.00 | 42.87±4.08 | 123658.03±24451.39 | 2.34±0.29 |
| #25 | 0.36 | 0.72 | 0.08 | 0.06 | 0.06172 | 98.74±7.44 | 43.07±4.07 | 123761.43±24124.14 | 2.31±0.28 |
| #26 | 0.36 | 0.7 | 0.08 | 0.06 | 0.06172 | 97.74±8.24 | 43.23±5.06 | 123762.59±28918.07 | 2.29±0.32 |
| #27 | 0.36 | 0.71 | 0.08 | 0.06 | 0.06172 | 98.00±8.13 | 43.99±4.00 | 127228.04±23587.83 | 2.25±0.29 |
| #28 | 0.36 | 0.69 | 0.08 | 0.06 | 0.06172 | 97.95±8.59 | 44.84±4.14 | 131678.86±25344.88 | 2.20±0.29 |
| #29 | 0.36 | 0.74 | 0.08 | 0.06 | 0.06172 | 99.88±8.41 | 44.87±4.22 | 135042.17±25925.36 | 2.25±0.29 |
| #30 | 0.36 | 0.68 | 0.08 | 0.06 | 0.06172 | 98.74±8.48 | 45.77±3.99 | 138018.62±25429.68 | 2.17±0.27 |
| #31 | 0.28 | 0.48 | 0.048 | 0.084 | 0.06172 | 97.13±8.80 | 50.05±5.20 | 160279.17±37446.26 | 1.96±0.24 |
| #32 | 0.36 | 0.66 | 0.08 | 0.06 | 0.06172 | 101.99±8.09 | 50.76±4.29 | 173264.93±31001.69 | 2.02±0.22 |
| #33 | 0.28 | 0.48 | 0.016 | 0.084 | 0.04936 | 101.65±6.74 | 51.54±3.12 | 176677.89±23753.79 | 1.98±0.17 |
| #34 | 0.36 | 0.72 | 0.008 | 0.06 | 0.06172 | 107.42±7.36 | 53.32±4.21 | 200909.70±32066.71 | 2.03±0.22 |
| #35 | 0.28 | 0.48 | 0.096 | 0.084 | 0.06172 | 107.06±9.28 | 56.34±5.59 | 222518.42±48898.68 | 1.92±0.22 |
| #36 | 0.28 | 0.48 | 0.008 | 0.084 | 0.04936 | 112.17±10.29 | 57.56±4.28 | 242863.59±41993.11 | 1.96±0.23 |
| #37 | 0.28 | 0.48 | 0.008 | 0.084 | 0.06172 | 111.78±9.90 | 60.10±4.54 | 261926.79±47182.96 | 1.87±0.20 |
| #38 | 0.36 | 0.72 | 0.008 | 0.06 | 0.06172 | 115.45±8.29 | 59.22±5.02 | 264887.68±46265.14 | 1.96±0.22 |
| #39 | 0.28 | 0.48 | 0.064 | 0.084 | 0.06172 | 113.81±9.23 | 61.90±5.58 | 283171.64±57169.27 | 1.85±0.19 |
| #40 | 0.36 | 0.2 | 0.08 | 0.06 | 0.06172 | 113.75±11.99 | 73.22±8.79 | 383387.23±105908.03 | 1.57±0.20 |
| #41 | 0.36 | 0.72 | 0.004 | 0.06 | 0.06172 | 126.62±8.24 | 74.85±4.95 | 448231.62±61762.56 | 1.70±0.16 |
| #42 | 0.28 | 0.48 | 0.032 | 0.084 | 0.06172 | 127.74±9.87 | 75.45±6.00 | 462263.15±84153.17 | 1.70±0.16 |
| #43 | 0.28 | 0.48 | 0.008 | 0.084 | 0.06172 | 131.58±9.68 | 78.00±5.71 | 507412.31±84549.05 | 1.69±0.15 |
| #44 | 0.28 | 0.72 | 0.008 | 0.06 | 0.04936 | 152.00±9.69 | 92.83±6.63 | 824252.22±128922.17 | 1.64±0.13 |

**Table.S1. Synthesis details and dimensions of gold nanorods.** This table summarizes the dimension of gold nanorods characterized by SEM images and the amount of chemicals used for the gold nanorod synthesis.